\documentclass[12pt]{article}
\usepackage{amsmath, amssymb}
\usepackage{mathrsfs}
\begin{document}
\date{}
\title{$su(1,1)$ algebraic approach of the Dirac equation with Coulomb-type scalar and vector potentials in $D+1$
dimensions}

\author{M. Salazar-Ram\'{\i}rez$^{a}$, D. Mart\'{\i}nez$^{b}$,\footnote{{\it E-mail address:} dmartinezs77@yahoo.com.mx}\\ R. D. Mota$^{c}$ and V. D. Granados$^{a}$} \maketitle

\begin{minipage}{0.9\textwidth}

\small $^{a}$ Escuela Superior de F\'{\i}sica y Matem\'aticas,
Instituto Polit\'ecnico Nacional,
Ed. 9, Unidad Profesional Adolfo L\'opez Mateos, 07738 M\'exico D F, M\'exico.\\

\small $^{b}$ Universidad Aut\'onoma de la Ciudad de M\'exico,
Plantel Cuautepec, Av. La Corona 320, Col. Loma la Palma,
Delegaci\'on
Gustavo A. Madero, 07160, M\'exico D. F., M\'exico.\\

\small $^{c}$ Unidad Profesional Interdisciplinaria de
Ingenier\'{\i}a y Tecnolog\'{\i}as Avanzadas, IPN. Av. Instituto
Polit\'ecnico Nacional 2580, Col. La Laguna Ticom\'an, Delegaci\'on
Gustavo A.
Madero, 07340 M\'exico D. F., M\'exico.\\

\end{minipage}

\begin{abstract}
We study the Dirac equation with Coulomb-type vector and scalar
potentials in $D+1$ dimensions from an $su(1,1)$ algebraic approach.
The generators of this algebra are constructed by using the
Schr\"odinger factorization. The theory of unitary representations
for the $su(1,1)$ Lie algebra allows us to obtain the energy
spectrum and the supersymmetric ground state. For the cases where
there exists either  scalar or vector potential our results are
reduced to those obtained by analytical techniques.
\end{abstract}

\section{Introduction}
As it is well known the factorization methods are very important to
study quantum systems \cite{DIRLIB,SCH1A,INF1}, since they are the basis for obtaining
the energy spectrum and  the eigenfunctions  in  an algebraic way \cite{INF1}.

The fundamental ideas of factorization in quantum physics were
settled by Dirac \cite{DIRLIB} and Schr\"odinger \cite{SCH1A}. However,
Infeld and Hull \cite{INF1} were the first to introduce a
systematic method to factorize and classify a large class of
potentials. Moreover, for several problems it has been shown that
factorization operators are directly related to the supersymmetric
charges \cite{COP,DUT,BAG,LAHI} introduced by Witten \cite{WIT}.

On the other hand, constants of motion for a given physical problem
allow to simplify the corresponding Hamiltonian by reducing its
degrees of freedom. Moreover, such conserved quantities are directly
associated with symmetry groups and compact and non-compact Lie
algebras \cite{WYBOURNE}. Such algebras play a central role to study
many properties of quantum systems because, among other things, they
are the basis for selection rules that forbid the existence of
certain states and processes. There is not a unified way to obtain
the explicit form of the generators of compact and non-compact
algebras, but several to find them, as it is shown in
\cite{BARUT,ENGLEFIELD}. However, in references \cite{MILLER,LEVAI,ASIM,PLA} attempts to
systematize the construction of the $su(1,1)$ Lie algebra generators
from factorization methods have been reported. Recently, in a series
of papers it has been shown that the Schr\"odinger factorization
operators can be used to construct the $su(1,1)$ algebra generators
for relativistic and non-relativistic central potential Hamiltonians
\cite{KeCo,NDIM,DANIEL}. These results enhance the importance of factorization
methods in solving quantum systems.

The three dimensional relativistic Kepler-Coulomb potential is one
of the solvable problems in physics. This is due to the conservation
of the total angular momentum and the Dirac and Lippmann-Johnson
operators \cite{DAHL}, which reduces to the Runge-Lenz vector in the
non-relativistic limit. The first two are due to the existence of
spin, and the latter gives account of the degeneracy in the
eigenvalues of the Dirac operator of the energy spectrum. Symmetries
and SUSY QM are intimately related. In fact, it has been shown that
supersymmetry is generated by the Lippmann-Johnson operator
\cite{DAHL}. In reference  \cite{KeCo} we have studied the
relativistic Kepler-Coulomb problem from an algebraic approach by
using the Schr\"odinger factorization to construct the $su(1,1)$
algebra generators. This problem admits to be treated in other ways:
analytical \cite{THALL1,THALL2,ROBIN1,ROBIN2,CHINOS1,CHINOS2} and factorization methods \cite{LIMA,SUKU,JARVIS},
shape-invariance \cite{LIMA}, SUSY QM for the first-
\cite{THALL1,SUKU} and second-order differential equations
\cite{JARVIS}, two-variable realizations of the $su(1,1)$ Lie
algebra \cite{MEX3} and using the Biedenharn-Temple operator
\cite{HORVATHY}.

In \cite{JOS} Joseph studied the  $1/r$ potential in  $D+1$
dimensions by means of the self-adjoint operators. Recently,  the
energy spectrum and the eigenfunctions of this problem were obtained
by solving the confluent hypergeometric equation \cite{YJIA,SHI1}.
Moreover, in \cite{JAPON} the Johnson-Lippmann operator for this
potential has been constructed and used to generate the SUSY
charges.

The Dirac equation with Coulomb-type vector and scalar potentials in
three dimensions has been solved by using SUSY QM \cite{JUG} and by an analytical approach \cite{GREINER2}. In
\cite{SHI2}, this problem has been solved in $D$-dimensions by
constructing the angular eigenfunctions from the group theory
and the radial equations solutions were expressed in terms of confluent
hypergeometric functions.

In the present Letter, we study the Dirac equation with Coulomb-type
vector and scalar potentials in $D+1$ dimensions from an $su(1,1)$
algebraic approach. We obtain the uncoupled second-order differential
equations for bound states satisfied by the radial components and by
applying the Schr\"odinger factorization we construct the corresponding $su(1,1)$ algebra generators. We use the theory of
unitary representations to obtain the energy spectrum and the action of the $su(1,1)$ algebra
generators on the radial eigenstates. Also, we find the Schr\"odinger and SUSY ground states. By particularizing our results to the cases where there exists either scalar or vector potential, or where both potentials are equal, we show that our
treatment successfully reproduce the results obtained by analytical
techniques. Finally, we give some concluding remarks.

\section{The relativistic Dirac equation in $D+1$ dimensions}

The Dirac equation in general dimensions is
\begin{equation}
H\Psi\equiv\left(\sum^D_{a=1}\alpha_a{p}_a+\beta\left(m+V_s\left(r\right)\right)+V_v\left(r\right)\right)\Psi=i\frac{\partial\Psi}{\partial{t}},
\end{equation}
with $\hbar=c=1$, $m$ is the mass of the particle, $p_a=-i\partial_a=-i\frac{\partial}{\partial{x}_a}$, $1\leq a\leq D$, $V_s$ and $V_v$ are the
spherically symmetric scalar and vector potentials, respectively. Moreover, $\alpha_a$ and $\beta$ satisfy the relations $\alpha_a\alpha_b+\alpha_b\alpha_a=2\delta_{ab}\textbf{1}$, $\alpha_a\beta+\beta\alpha_a=0$ and $\alpha_a^2=\beta^2=\texttt{1}$ \cite{SHI2}. In $D$ spatial dimensions the orbital angular momentum operators $L_{ab}$, the spinor operators $S_{ab}$ and the total angular
momentum operators $J_{ab}$ are defined as $L_{ab}=-L_{ba}=ix_a\frac{\partial}{\partial
x_b}-ix_b\frac{\partial}{\partial x_a}$, $S_{ab}=-S_{ba}=i\frac{\alpha_a\alpha_b}{2}$ and $J_{ab}=L_{ab}+S_{ab}$, respectively. Thus,
$L^2=\sum_{a<b}^D L^2_{ab}$, $S^2=\sum_{a<b}^D S^2_{ab}$, $J^2=\sum_{a<b}^D J^2_{ab}$, with $1\leq a\leq b\leq D$. Hence, for a spherically symmetric potential, the total angular
momentum operator $J_{ab}$ and the spin-orbit operator
$K_D=-\beta\left(J^2-L^2-S^2+\frac{\left(D-1\right)}{2}\right)$
commute with the Dirac Hamiltonian, $H$. For a given total angular
momentum $j$, the eigenvalues of $K_D$ are
$\kappa_D=\pm\left(j+\left(D-2\right)/2\right)$, where the minus
sign is for aligned spin $j=\ell+\frac{1}{2}$, and the plus sign is
for unaligned spin $\ell-\frac{1}{2}$ \cite{SHI2}.

We propose the wave function of the Dirac Hamiltonian $H$ to be of
the form
\begin{equation}\label{SolDir}
\Psi(\vec r, t)=r^{-\frac{D-1}{2}}
\begin{pmatrix}
G_{\kappa_D}^{(1)}\left(r\right)Y_{jm}^\ell(\Omega_D)\\
iG_{\kappa_D}^{(2)}\left(r\right)Y_{jm}^{\ell'}(\Omega_D)
\end{pmatrix}e^{-iEt},
\end{equation}
being $G^{(1,2)}_{k_D}(r)$ the radial functions,
$Y_{jm}^\ell(\Omega_D)$ and $Y_{jm}^{\ell'}(\Omega_D)$ the
hyperspherical harmonic functions coupled with the total angular
momentum quantum number $j$ and $E$ the energy. Thus, the Dirac equation leads to the radial equation
\begin{equation}\label{difGG}
\begin{pmatrix}
\frac{dG^{(1)}_{\kappa_D}}{dr}\\
\frac{dG^{(2)}_{\kappa_D}}{dr}
\end{pmatrix}=
\begin{pmatrix}
-\frac{\kappa_D}{r} & V_s-V_v+m+E \\
V_v-V_s+m-E & \frac{\kappa_D}{r} \end{pmatrix}
\begin{pmatrix}
G^{(1)}_{\kappa_D}\\
G^{(2)}_{\kappa_D}
\end{pmatrix}.
\end{equation}

In this work, we consider the Coulomb-type scalar and vector
potentials $V_v= -\frac{\alpha}{r}$, $V_s=-\frac{\alpha'}{r}$,
with $\alpha$ and $\alpha'$ positive constants. In the following sections we study separately the cases $\alpha\neq\alpha'$ and $\alpha=\alpha'$.

\section{Case $\alpha\neq\alpha'$}

By introducing the new variable
\begin{equation}\label{rho}
\rho = \left(\alpha E
+\alpha{'}m\right)r/\sqrt{|{\alpha'}^2-\alpha^2|},
\end{equation}
equation (\ref{difGG}) can be written as
\begin{align}\label{acopla1}
\textstyle
A^{-}F^{(1)}_{\kappa_D}&=-\textstyle\frac{1}{\gamma}\sqrt{\frac{|{\alpha'}-\alpha|}{\alpha+\alpha{'}}}\left(\kappa_D+\gamma\frac{\alpha{'}E+\alpha{m}}{\alpha{E}+\alpha{'}m}\right)F^{(2)}_{\kappa_D},\\\label{acopla2}
\textstyle
A^{+}F^{(2)}_{\kappa_D}&=-\textstyle{sgn}(\alpha-\alpha')\frac{1}{\gamma}\sqrt{\frac{\alpha+\alpha{'}}{|{\alpha'}-\alpha|}}\left(\kappa_D-\gamma\frac{\alpha{'}E+\alpha{m}}{\alpha{E}+\alpha{'}m}\right)F^{(1)}_{\kappa_D},
\end{align}
where
\begin{equation}
A^{\pm}=\pm\frac{d}{d\rho}-\frac{\gamma}{\rho}+\frac{\sqrt{|\alpha'^2-\alpha^2|}}{\gamma},
\end{equation}
\begin{equation}
\begin{pmatrix}
F^{(1)}_{\kappa_D}\\
F^{(2)}_{\kappa_D}
\end{pmatrix}= \begin{pmatrix}
\kappa_D+\gamma&-\left(\alpha-\alpha{'}\right)\\
-\left(\alpha+\alpha{'}\right)&
\kappa_D+\gamma\end{pmatrix}\begin{pmatrix}
G^{(1)}_{\kappa_D}\\
G^{(2)}_{\kappa_D}
\end{pmatrix},
\end{equation}
\begin{equation}\label{sgn}
sgn(\alpha-\alpha')=\ \begin{cases}
 \hspace{0.3cm} 1 \hspace{0.6cm}  \text{if}\quad \alpha>\alpha'\\
-1 \hspace{0.6cm}\text{if}\quad \alpha<\alpha'
\end{cases}
\end{equation}
and $\gamma=\sqrt{\kappa_D^2+\alpha{'}^2-\alpha^2}$.

From equations (\ref{acopla1}) and (\ref{acopla2}) we obtain the
uncoupled second-order differential equations
\begin{align}
\label{desacop1}
\left(-\frac{d^2}{d\rho^2} +
\frac{\gamma\left(\gamma\pm1\right)}{\rho^2}+\frac{|{\alpha'}^2-\alpha^2|}{\gamma^2}-\frac{2\sqrt{|{\alpha'}^2-\alpha^2|}}{\rho}\right)F^{(1,2)}_{\kappa_D}=\nonumber\\
sgn\left(\alpha-\alpha'\right)\frac{1}{\gamma^2}\left(\kappa_D^2-\gamma^2\left(\frac{\alpha{'}E+\alpha{m}}{\alpha{E}+\alpha{'}m}\right)^2\right)F^{(1,2)}_{\kappa_D}
\end{align}
where the superscripts $(1,2)$ correspond to $(+,-)$ in the centrifugal
term. Thus, for bound states $(m^2>E^2)$, equation (\ref{desacop1}) is written as
\begin{align}
\left(-\rho^2\frac{d^2}{d\rho^2}+\xi^2\rho^2-2\sqrt{|{\alpha'}^2-\alpha^2|}\rho\right)F^{(1)}_{\kappa_D}& =-\gamma\left(\gamma+1\right)F^{(1)}_{\kappa_D},\label{desacopla1}\\
\left(-\rho^2\frac{d^2}{d\rho^2}+\xi^2\rho^2-2\sqrt{|{\alpha'}^2-\alpha^2|}\rho\right)F^{(2)}_{\kappa_D}&
=-\gamma\left(\gamma-1\right)F^{(2)}_{\kappa_D},\label{desacopla2}
\end{align}
with
\begin{equation}\label{xiD}
\xi^2=sgn(\alpha-\alpha')\left[\left(\frac{\alpha{'}E+\alpha{m}}{\alpha{E}+\alpha{'}m}\right)^2-1\right].
\end{equation}
It must be noted that equation (\ref{desacopla1}) is formally
obtained from equation (\ref{desacopla2}) by performing the change
$\gamma\rightarrow \gamma+1$. Hence, by defining
$\psi_{\gamma}\equiv F^{(2)}_{\kappa_D}$, it implies that $F^{(1)}_{\kappa_{\scriptstyle D}}\propto\psi_{\gamma+1}$. Thus, the solution for the Dirac equation in spinorial form can be
expressed as $\Phi_{\kappa_D}\equiv\begin{pmatrix}F^{(1)}_{\kappa_D}\\F^{(2)}_{\kappa_D}\end{pmatrix}=\begin{pmatrix}\psi_{\gamma+1}\\\psi_{\gamma}\end{pmatrix}$.

\subsection{The $su(1,1)$ Lie algebra and the energy spectrum}

In order to construct the $su(1,1)$ algebra generators, we apply the
Schr\"odinger factorization \cite{SCH1A,DANIEL} to the left-hand
side of equation (\ref{desacopla2}). Thus, we propose 
\begin{equation}
\left(\rho\frac{d}{d\rho}+a\rho+b\right)\left(-\rho\frac{d}{d\rho}+c\rho+f\right)\psi_{\gamma}=g\psi_{\gamma},
\label{sch}
\end{equation}
where $a$, $b$, $c$, $f$ and $g$ are constants to be determined.

Expanding this expressions and comparing it with equation
(\ref{desacopla2}) we obtain $a=c=\pm\xi$, $f=1+b=\mp\frac{\sqrt{|{\alpha'}^2-\alpha^2|}}{\xi}$ and $g=b(b+1)-\gamma(\gamma-1)$. Therefore, the differential equation satisfied by the lower
component of the spinor $\Phi_{k_D}$ is factorized as
\begin{align}
\left(\mathscr{L}_\mp\mp1\right)\mathscr{L}_\pm\psi_{\gamma}
=\left[\left(\frac{\sqrt{|{\alpha'}^2-\alpha^2|}}{\xi}\pm\frac{1}{2}\right)^2-\left(\gamma-\frac{1}{2}\right)^2\right]\psi_{\gamma},\label{Facdesc1}
\end{align}
where
\begin{equation}\label{OpeFac}
\mathscr{L}_\pm =\mp\rho\frac{d}{d\rho}
+\xi\rho-\frac{\sqrt{|{\alpha'}^2-\alpha^2|}}{\xi}
\end{equation}
are the Schr\"odinger operators.

Also, we define the new pair of operators
\begin{align}
\Pi_\pm &
\equiv\mp\rho\frac{d}{d\rho}+\xi\rho-\Pi_3,\label{Tmasmen1}
\end{align}
where
\begin{align}
\Pi_3\psi_{\gamma}\equiv\frac{1}{2\xi}\left(-\rho\frac{d^2}{d\rho^2}
+\xi^2\rho+\frac{\gamma\left(\gamma-1\right)}{\rho}\right)\psi_{\gamma}=\frac{\sqrt{|{\alpha'}^2-\alpha^2|}}{\xi}\psi_{\gamma}\label{T3dif2or1}
\end{align}
has been obtained from equation (\ref{desacopla2}).

A direct calculation shows that operators $\Pi_\pm$ and $\Pi_3$
close the $su(1,1)$ Lie algebra
\begin{equation}
\left[\Pi_\pm,\Pi_3\right]=\mp\Pi_\pm,\hspace{3ex}\left[\Pi_+,\Pi_-\right]=-2\Pi_3.\label{Relcom21}
\end{equation}

Therefore, from equations  (\ref{Tmasmen1}) and (\ref{T3dif2or1})
we find that the corresponding quadratic Casimir operator $\Pi^2=-\Pi_{+}\Pi_{-}+\left(\Pi_3\right)^2 - \Pi_3$, satisfies the eigenvalue equation
\begin{align}
\Pi^2\psi_{\gamma}=\gamma\left(\gamma-1\right)\psi_{\gamma}.\label{CasEn1}
\end{align}

As we mentioned above, by performing the change $\gamma\rightarrow
\gamma+1$ to $\psi_{\gamma}$, we obtain the upper component of the
spinor $\Phi_{k_D}$, $\psi_{\gamma+1}$. Thus, the $su(1,1)$ algebra
generators for this component are
\begin{align}
\Gamma_3&\equiv\frac{1}{2\xi}\left(-\rho\frac{d^2}{d\rho^2}
+\xi^2\rho+\frac{\gamma\left(\gamma+1\right)}{\rho}\right),\label{T3dif2or2}\\
\Gamma_\pm &
\equiv\mp\rho\frac{d}{d\rho}+\xi\rho-\Gamma_3.\label{Tmasmen2}
\end{align}

In order to determine the properties of the operators $\Pi_3$ and
$\Pi_\pm$, we introduce the inner product on the Hilbert space
spanned by the radial eigenfunctions for the Dirac equation with scalar and vector potentials in $D$ dimensions \cite{ADAMS}
\begin{equation}
\left(\phi,\zeta\right)\equiv\int_0^\infty
\phi^{*}\left(\rho\right)\zeta\left(\rho\right)\rho^{-1}d\rho.\label{ortonor}
\end{equation}

Thus, the operator $\Pi_3$ is Hermitian with respect to this scalar
product. Moreover, using equations (\ref{Tmasmen1}) and
(\ref{ortonor}), it can be proved that operators $\Pi_\pm$ are
hermitian conjugates, $\Pi_\pm=\Pi_\mp^\dagger$.

The theory of unitary irreducible representations of the $su(1,1)$ Lie algebra has been studied in several works
\cite{ADAMS,ADAMS1} and it is based on the equations
\begin{align}
T^2\vert\mu\:\nu\rangle& =\mu(\mu+1)\vert\mu\:\nu\rangle,\label{C2}\\
T_3\vert\mu\:\nu\rangle& =\nu\vert\mu\:\nu\rangle\label{C3},\\
T_\pm\vert\mu\:\nu\rangle&
=\left[(\nu\mp\mu)(\nu\pm\mu\pm1)\right]^{1/2}\vert\mu\:\nu\pm1\rangle\label{T+-},
\end{align}
where $T^2$ is the quadratic Casimir operator, $\nu=\mu+q+1$,
$q=0,1,2,...$ and $\mu>-1$.  From equation (\ref{T+-}) it can be
noted that $T_+$($T_-$) is the raising (lowering) operator for
$\nu$.

Thus, from equations (\ref{CasEn1}) and (\ref{C2}), and
(\ref{T3dif2or1}) and (\ref{C3}), we find that
\begin{align}
\mu_{\gamma}& =\gamma-1,\label{mu1}\\
\nu_{\gamma}&
=n_{\gamma}+\gamma=\frac{\sqrt{|{\alpha'}^2-\alpha^2|}}{\xi},\label{nu1}
\end{align}
respectively, with $n_{\gamma}=0,1,2,...$.

Since the operators $\Pi_\pm$ leave fixed the quantum number
$\mu_{\gamma}$, equation (\ref{mu1}) ensures that the values of
$\gamma$ remain fixed. This is because the change
$\nu_{\gamma}\rightarrow\nu_{\gamma}\pm1$ induced by the operators
$\Pi_\pm$ on the basis vectors
$\vert\mu_{\gamma}\:\nu_{\gamma}\rangle$ implies
$n_{\gamma}\rightarrow n_{\gamma}\pm1$. Thus, by setting
$\vert\mu_{\gamma}\:\nu_{\gamma}\rangle\rightarrow\psi_{\gamma}^{n_{\gamma}}$
and from equations (\ref{T+-}), (\ref{mu1}) and (\ref{nu1}), we find
that
\begin{align}
\Pi_{\pm}\psi_{\gamma}^{n_{\gamma}}=C_{\pm}^{\left(n_{\gamma},\gamma\right)}\psi_{\gamma}^{n_{\gamma}\pm1},\label{Pi+-}
\end{align}
with
$C_{\pm}^{\left(n_{\gamma},\gamma\right)}=\left[\left(n_{\gamma}+\gamma\mp
\gamma\pm 1\right)\left(n_{\gamma}+\gamma\pm
\gamma\right)\right]^{1/2}$ a real number.

By using the equations (\ref{xiD}) and (\ref{nu1}) we find that the
energy spectrum for the lower component of the spinor $\Phi_{\kappa_D}$, $\psi_{\gamma}^{n_{\gamma}}$, is
\begin{equation}
E_{\gamma}/m=\left[-\alpha{'}\alpha\pm\zeta\sqrt{\zeta^2+\alpha^2-(\alpha')^2}\right]/\left(\alpha^2+\zeta^2\right),
\end{equation}
where $\zeta\equiv\gamma+n_{\gamma}$.

If we perform the change $\gamma\rightarrow\gamma+1$ in equation
(\ref{Pi+-}), we find that the action of the ladder operators
$\Gamma_\pm$ on the functions $\psi_{\gamma+1}^{n_{\gamma+1}}$ is
\begin{align}
\Gamma_{\pm}\psi_{\gamma+1}^{n_{\gamma+1}}=C_{\pm}^{\left(n_{\gamma+1},\gamma+1\right)}\psi_{\gamma+1}^{n_{\gamma+1}\pm1},\label{Ga+-}
\end{align}
meanwhile the energy spectrum for the upper component of the spinor $\Phi_{\kappa_D}$ is
\begin{equation}
E_{\gamma+1}/m=\left[-\alpha{'}\alpha\pm\zeta'\sqrt{(\zeta')^2+\alpha^2-(\alpha')^2}\right]/\left(\alpha^2+(\zeta')^2\right),
\end{equation}
where $\zeta'\equiv\gamma+n_{\gamma+1}+1$. However, since $\psi_{\gamma+1}^{n_{\gamma+1}}$ and
$\psi_{\gamma}^{n_{\gamma}}$ are the components of the same spinor, they should have the
same energy, $E_{\gamma}=E_{\gamma+1}$. Thus, we obtain $n\equiv n_{\gamma}=n_{\gamma+1}+1$, where $n=0,1,2,3,...$ is the radial quantum number. 

Therefore, the energy spectrum for the Dirac Hamiltonian with scalar and vector
potentials in $D$ dimensions can be written as
\begin{equation}
\label{EnerFinG}
E=
m\left[\frac{-\alpha{'}\alpha}{\alpha^2+\left(\gamma+n\right)^2}\pm\sqrt{\left(\frac{\alpha{'}\alpha}{\alpha^2+\left(\gamma+n\right)^2}\right)^2-\left(\frac{\alpha{'}^2-\left(\gamma+n\right)^2}{\alpha^2+\left(\gamma+n\right)^2}
\right)}\right]
\end{equation}
which is the same reported in \cite{SHI2}, obtained from an
analytical point of view. Moreover, this expression is reduced to that for the three dimensional case \cite{GREINER2}.

In this way we have shown that
$\Phi_{\kappa_D}^n$ is given by
\begin{equation}
\Phi_{\kappa_D}^n=\begin{pmatrix}
F_{n\kappa_D}^{(1)}\\
F_{n\kappa_D}^{(2)}
\end{pmatrix}=\begin{pmatrix}
\psi_{\gamma+1}^{n-1}\\
\psi_{\gamma}^n
\end{pmatrix}.\label{2spinor}
\end{equation}
The relation between the components of the spinor above via the
radial quantum number $n$, deduced from the theory of unitary
representation can be observed from an analytical approach, as it is
shown below.

\subsection{The Schr\"odinger and SUSY QM ground states}

From equation (\ref{Pi+-}), for $n=0$ we find that
only the state $\psi_{\gamma\;SCH}^{0}=\eta\rho^{\gamma}
e^{-\sqrt{|\alpha'^2-\alpha^2|}\rho/\gamma}$ is normalizable with respect to the inner product defined in
expression (\ref{ortonor}) and satisfies the differential equation
$\Pi_-\psi_{\gamma}^{0}=0$. Moreover, for $n=0$ and from the definition of the radial quantum number, we find that $n_{\gamma+1}=-1$. For this value,
$C_{-}^{\left(-1,\gamma+1\right)}$ results in a complex number. From
the theory of unitary representation and equation (\ref{Ga+-}), the
function $\psi_{\gamma+1}^{-1}$ is non-normalizable \cite{ADAMS1}.
Therefore, this function is not a physically acceptable solution and
the spinor corresponding to $n=0$ is
\begin{equation}
\Phi_{\gamma\;\text{SCH}}^0=\begin{pmatrix}
0\\
\rho^{\gamma} e^{-\sqrt{|\alpha'^2-\alpha^2|}\rho/\gamma}
\end{pmatrix},\label{spinor1}
\end{equation}
which we denote as the Schr\"odinger ground state.

In fact, the operators $A^{\pm}$ in equations (\ref{acopla1}) and (\ref{acopla2})
are the SUSY operators, from which the partner Hamiltonians are given by $H_+=A^-A^+$ and $H_-=A^+A^-$. The ground state for $H_-$ satisfies the condition $A^-\psi_{\gamma\;\text{SUSY}}^{0}=0$, that leads to the
square-integrable eigenfunction $\psi_{\gamma\;SUSY}^{0}=c\rho^\gamma
\exp\left(-\rho\sqrt{|\alpha'^2-\alpha^2|}/\gamma\right)$. On the other hand, the solution for the equation $A^+\psi_{\gamma+1\;SUSY}^{0}=0$ is $\psi_{\gamma+1\;\text{SUSY}}^{0} =c\rho^{-\gamma}
\exp\left(-\rho\sqrt{|\alpha'^2-\alpha^2|}/\gamma\right)$. Since this function is not a square-integrable then, it is not a
physically acceptable solution and it must be taken as the zero
function. Hence, the SUSY ground-state is given by
\begin{equation}
\Phi_{\gamma\;\text{SUSY}}^0=\begin{pmatrix}
0\\
\rho^{\gamma} e^{-\sqrt{|\alpha'^2-\alpha^2|}\rho/\gamma}
\end{pmatrix}\label{spinor2}.
\end{equation}
The above results leads us to the conclusion that the Schr\"odinger
and SUSY ground states for the relativistic Dirac equation with
Coulomb-type scalar and vector potentials are the same.

The explicit form of $\Phi_{k_D}^n$ corresponding to higher energy
levels can be obtained from an analytical approach. In order to
solve the differential equations (\ref{desacopla2}) we propose $F_{\kappa_D}^{(2)}=\rho^{\gamma}e^{-\xi\rho} f(\rho)$, where $f(\rho)$ must satisfy
\begin{equation}
\left[y\frac{d^2}{dy^2}+(2\gamma-y)\frac{d}{dy}+\frac{\sqrt{|{\alpha'}^2-\alpha^2|}}{\xi}-\gamma\right]
f\left(y/2\xi\right)=0,
\end{equation}
with $y=2\xi\rho$. The solution for this equation is the confluent
hypergeometric function $_1F_1\left(-n,2\gamma;2\xi\rho\right)$.
Thus, we get $F_{n\kappa_D}^{(2)}=\eta_2\rho^{\gamma}{e}^{-\xi\rho}\;_1F_1\left(-n,2\gamma;2\xi\rho\right)$. In a similar way, the solution for equation (\ref{desacopla1}) is $F_{n\kappa_D}^{(1)}=\eta_1\rho^{\gamma+1}e^{-\xi\rho}\;$ $_1F_1\left(-n+1,2\gamma+2;2\xi\rho\right)$. The relation between these solutions is in agreement with that obtained from the theory
of unitary representations, equation (\ref{2spinor}). Moreover, it
is known that $_1F_1(0,b;z)=1$, whereas $_1F_1(a,b;z)$ diverges for
$a>0$. Therefore, for $n=0$, $F_{0\kappa_D}^{(2)}=\eta_{\gamma}\rho^{\gamma}{e}^{-\xi\rho}$, while the function $F_{0\kappa_D}^{(1)}$ is not square-integrable, and it must be taken as the zero function. This result is in
accordance with equation (\ref{spinor1}), which has been obtained
from an algebraic approach.

From the definition of $n$, equations (\ref{Pi+-}) and
(\ref{Ga+-}) imply
\begin{equation}
\Pi_\pm\psi_{\gamma}^n\propto\psi_{\gamma}^{n\pm1},\hspace{3ex}\Gamma_\pm\psi_{\gamma+1}^{n-1}\propto
\psi_{\gamma+1}^{n-1\pm1}.\label{gs1}
\end{equation}

It is worth to notice that equation (\ref{gs1}) is valid for any
value of the radial quantum number except for $n=0$. This is because
the upper component of the spinor for $n=1$ can not be obtained from
the action of the operator $\Gamma_{+}$ on the upper component of
the Schr\"odinger ground state, equation (\ref{spinor1}).

Equations (\ref{T3dif2or1}) and (\ref{gs1}) allow us to
show that the action of the Schr\"odinger operators on the states
$\psi_{\gamma}^{n}$ is $\mathscr{L}_\pm\psi_{\gamma}^{n}\propto\psi_{\gamma}^{n\pm1}$. A similar result can be obtained from the
operators $\Gamma_\pm$. Therefore, we conclude that the action of the Schr\"odinger operators
$\mathscr{L}_\pm$ on the components of the spinor
$\Phi_{\kappa_D}^n$ is to change only the radial quantum number $n$
leaving fixed the Dirac quantum number $\kappa_D=\kappa_D(\gamma)$.

Some particular cases of our general results obtained above are
shown in table 1.

\begin{table}
\begin{tabular}{|c|c|c|c|c|}
\hline
 $\text{Cases}$ &\hspace*{-0.3cm} $\text{Spatial}\hspace*{-0.1cm}$                     & \hspace*{-0.3cm}$\text{Potential}$ \hspace*{-0.3cm}                     & \hspace*{-0.3cm}$\gamma$\hspace*{-0.3cm}& $\xi^2$  \\
 $            $ &\hspace*{-0.3cm} $\text{Dimensions}\hspace*{-0.1cm}$                  &  \hspace*{-0.3cm}$\text{Parameters}$ \hspace*{-0.3cm}                                         &      &           \\
\hline
$I$ & $D$  & $\alpha\neq0,\alpha'=0$        & $\sqrt{\kappa_D^2-\alpha^2}$             &  $\left(\frac{m}{E}\right)^2-1$ \\
\hline
$II$ & $D$  & $\alpha'\neq0,\alpha=0$       & $\sqrt{\kappa_D^2+\alpha'^2}$            &  $1-\left(\frac{E}{m}\right)^2$ \\
\hline
$III$ & $3$  & $\alpha'\neq0, \alpha\neq0$  & $\sqrt{\kappa_3^2+\alpha'^2-\alpha^2}$   &  $sgn(\alpha-\alpha')\left[\left(\frac{\alpha{'}E+\alpha{m}}{\alpha{E}+\alpha{'}m}\right)^2-1\right]$ \\
\hline
$IV$ & $3$  & $\alpha\neq0,\alpha'=0$       &$\sqrt{\kappa_3^2-\alpha^2}$              &  $\left(\frac{m}{E}\right)^2-1$ \\
\hline
$V$ & $3$  & $\alpha'\neq0,\alpha=0$        & $\sqrt{\kappa_3^2+\alpha'^2}$            &  $1-\left(\frac{E}{m}\right)^2$ \\
\hline
\end{tabular}
\caption{\small It is shown the expressions for $\gamma$ and $\xi^2$ for
particular values of $\alpha$ and $\alpha'$ in $D$ or three spatial
dimensions.}
\end{table}

From an analytical point of view, case $I$ was studied in \cite{YJIA},
meanwhile cases $III$, $IV$ and $V$ were treated in \cite{GREINER2}.
On the other hand, case $III$ was studied by SUSY QM in \cite{JUG} and case $IV$ was treated algebraically
in \cite{KeCo}. It must be pointed out that our present results, which were obtained from an $su(1,1)$
algebraic approach, are in full agreement with those reported in the
last references.

\section{Case $\alpha=\alpha'$}

Because of the change of variable given in (\ref{rho}),
$\alpha=\alpha'$ is not an special case of the results of the
previous section. For the present case, from the Dirac radial equation (\ref{difGG}), we obtain the uncoupled second-order differential equations
\begin{align}
\left(-\rho^2\frac{d^2}{d\rho^2}+\xi^2\rho^2-2\alpha\rho\right)F^{(1)}_{\kappa_D}& =-\kappa_D\left(\kappa_D+1\right)F^{(1)}_{\kappa_D},\label{des1alpha=}\\
\left(-\rho^2\frac{d^2}{d\rho^2}+\xi^2\rho^2-2\alpha\rho\right)F^{(2)}_{\kappa_D}&
=-\kappa_D\left(\kappa_D-1\right)F^{(2)}_{\kappa_D},\label{desa2alpha=}
\end{align}
where $\xi^2=\frac{m-E}{m+E}$, $\rho=\left(m+E\right)r$ and
\begin{equation}
\begin{pmatrix}
F^{(1)}_{\kappa_D}\\
F^{(2)}_{\kappa_D}
\end{pmatrix}= \begin{pmatrix}
\kappa_D&0\\
-\alpha & \kappa_D\end{pmatrix}\begin{pmatrix}
G^{(1)}_{\kappa_D}\\
G^{(2)}_{\kappa_D}
\end{pmatrix}.
\end{equation}

Since equation (\ref{des1alpha=}) is formally
obtained from equation (\ref{desa2alpha=}) by performing the change
$\kappa_D\rightarrow \kappa_D+1$ then, by defining
$\psi_{\kappa_D}\equiv F^{(2)}_{\kappa_D}$, we conclude that the solution to the Dirac equation in spinorial form is
\begin{equation}
\Phi_{\kappa_D}\equiv\begin{pmatrix}
F^{(1)}_{\kappa_D}\\
F^{(2)}_{\kappa_D}
\end{pmatrix}=
\begin{pmatrix}
\psi_{\kappa_D+1}\\
\psi_{\kappa_D}
\end{pmatrix}.\label{spinoralpha=}
\end{equation}

Proceeding in a similar way as in the case $\alpha\neq\alpha'$, we obtain that the
$su(1,1)$ algebra generators for the upper and lower components of
$\Phi_{\kappa_D}$ are
\begin{align}
\chi_3 &=\chi_3(\kappa_D)\equiv\frac{1}{2\xi}\left(-\rho\frac{d^2}{d\rho^2}
+\xi^2\rho+\frac{\kappa_D\left(\kappa_D-1\right)}{\rho}\right),\label{T3alpha=}\\
\chi_{\pm} &=\chi_\pm(\kappa_D)\equiv\mp\rho\frac{d}{d\rho}+\xi\rho-\chi_3\label{opemasmenalpha=}
\end{align}
and
\begin{equation}
\Lambda_3\equiv\chi_3(\kappa_D+1),\hspace{3ex}\label{T32alpha=}
\Lambda_\pm\equiv\chi_{\pm}(\kappa_D+1),
\end{equation}
respectively, with $\chi_3\psi_{\kappa_D}=(\alpha/\xi)\psi_{\kappa_D}$ and $\Lambda_3\psi_{\kappa_D+1}=(\alpha/\xi)\psi_{\kappa_D+1}$.

From the theory of unitary representations, equations (\ref{C2}),
(\ref{C3}) and (\ref{T+-}), we obtain that the corresponding energy
spectrum is
\begin{equation}
E= m\left[1-\frac{2\alpha^2}{\alpha^2+\left(n+|\kappa_D|\right)^2}
\right].\label{EnerFin}
\end{equation}
Also, we find that the Schr\"odinger and SUSY ground states are the
same and are given by
\begin{equation}
\Psi_{\text{SCH,SUSY}}^0=\begin{cases}
\begin{pmatrix}
0\\
\rho^{\kappa_D} e^{-\alpha\rho/\kappa_D}
\end{pmatrix}\hspace{0.8cm}\text{for}\hspace{0.4cm} \kappa_D>0,\vspace{0.2cm}\\
\begin{pmatrix}
\rho^{|\kappa_D|} e^{-\alpha\rho/|\kappa_D|}\\
0
\end{pmatrix}\hspace{0.5cm}\text{for}\hspace{0.4cm} \kappa_D<0.
\end{cases}
\end{equation}
These results are in accordance to those reported in \cite{JUG} for
three spatial dimensions.

\section{Concluding remarks}
We studied the radial Dirac equations with Coulomb-type scalar and
vector potentials in $D+1$ dimensions from an $su(1,1)$ algebraic
approach. With the Schr\"odinger factorization we were able to
construct two sets of $su(1,1)$ algebra generators. We applied the
theory of unitary representations for this algebra to find the
general form of the spinor wave function, the energy spectrum and
the SUSY ground state. We proved that Schr\"odinger and SUSY ground
states are the same. We showed that the action of the Schr\"odinger
operators on the radial eigenstates is to change only the radial
quantum number $n$ leaving fixed the Dirac quantum number
$\kappa_D$. The noncompactness of the $su(1,1)$ algebra reflects
that, for a fixed number $\kappa_D$, the quantum number $n$ is
bounded from below and unbounded from above. We tested our results
by matching different cases where there exists either scalar or
vector potentials in $D$ or three spatial dimensions
\cite{KeCo,YJIA,GREINER2,JUG,SHI2}. To our knowledge this is the first
time where the Dirac equation in $D$ dimensions with both scalar and
vector potential has been treated by an $su(1,1)$ algebraic approach.

Finally, our technique can be successfully applied to solve other relativistic problems like the Coulomb field with position-dependent mass \cite{ALH1} or the Dirac-Morse \cite{ALH2}, which are works in progress.

\section*{Acknowledgments}
This work was partially supported by SNI-M\'exico, COFAA-IPN,
EDI-IPN, SIP-IPN project number $20110127$ , and ADI-UACM project
number 7DA2023001.

\end{document}